\documentclass[conference]{IEEEtran}
\IEEEoverridecommandlockouts
\usepackage{cite}
\usepackage{amsmath,amssymb,amsfonts}
\usepackage{algpseudocode}
\usepackage{algorithm}
\usepackage{graphicx}
\usepackage{textcomp}
\usepackage{xcolor}
\usepackage{hyperref}
\usepackage[nolist]{acronym}

\begin{acronym}
    \acro{MIMO}{multiple-input, multiple-output}
    \acro{AWGN}{additive white Gaussian noise}
    \acro{CU}{central unit}
    \acro{KPI}{key performance indicator}
      \acro{KPIs}{key performance indicators}  
    \acro{D2D}{device-to-device}
    \acro{PAPR}{peak-to-average power ratio}
    \acro{ISAC}{integrated sensing and communication}
    \acro{DFRC}{dual-functional radar and communication}
    \acro{AoA}{angle of arrival}
    \acro{AoAs}{angles of arrival}
    \acro{ToA}{time of arrival}
    \acro{ToAs}{times of arrival}
    \acro{CFO}{carrier frequency offset}
    \acro{STO}{sampling time offset}
    \acro{RF}{radio frequency}
    \acro{URLLC}{ultra-reliable low latency communications}
    \acro{mMTC}{massive machine type communications}
	\acro{QCQP}{quadratic constrained quadratic programming}
	\acro{BnB}{branch and bound}
	\acro{SER}{symbol error rate}
	\acro{LFM}{linear frequency modulation}
	\acro{BS}{base station}
	\acro{UE}{user equipment}
	\acro{DL}{deep learning}
	\acro{CS}{compressed sensing}
	\acro{PR}{passive radar}
    \acro{PRs} {passive radars}
	\acro{LMS}{least mean squares}
	\acro{NLMS}{normalized least mean squares}
	\acro{RIS}{reconfigurable intelligent surface}
	\acro{RISs}{reconfigurable intelligent surfaces}
	\acro{IRS}{intelligent reflecting surface}
	\acro{IRSs}{intelligent reflecting surfaces}
	\acro{LISA}{large intelligent surface/antennas}
	\acro{LISs}{large intelligent surfaces}
	\acro{OFDM}{orthogonal frequency-division multiplexing}
	\acro{EM}{electromagnetic}
	\acro{MSE}{mean squared error}
	\acro{SNR}{signal-to-noise ratio}
	\acro{SRP}{successful recovery probability}
	\acro{CDF}{cumulative distribution function}
	\acro{ULA}{uniform linear array}
	\acro{RSS}{received signal strength}
	\acro{LoS}{line of sight}
	\acro{AoD}{angle of departure}
	\acro{AP}{access point}
	\acro{ML}{maximum likelihood}
	\acro{FIM}{Fisher information matrix}
    \acro{ZC}{Zadoff-Chu}
    \acro{CAZAC}{constant amplitude zero autocorrelation}
    \acro{DAC}{digital to analog converter}
    \acro{ADC}{analog to digital converter}
    \acro{UE}{user equipment}
    \acro{UEs}{user equipments}
\end{acronym}

\DeclareMathOperator{\SNR}{SNR}

\DeclareMathOperator{\dB}{dB}

\DeclareMathOperator{\MSE}{MSE}
\DeclareMathOperator{\SRP}{SRP}

\DeclareMathOperator{\diag}{diag}

\DeclareMathOperator*{\argsmax}{args\,max}

\def\BibTeX{{\rm B\kern-.05em{\sc i\kern-.025em b}\kern-.08em
    T\kern-.1667em\lower.7ex\hbox{E}\kern-.125emX}}
    \makeatletter
 \let\old@ps@headings\ps@headings
 \let\old@ps@IEEEtitlepagestyle\ps@IEEEtitlepagestyle
 \def\confheader#1{%
 \def\ps@headings{%
 \old@ps@headings%
 \def\@oddhead{\strut\hfill#1\hfill\strut}%
 \def\@evenhead{\strut\hfill#1\hfill\strut}%
 }%
 \def\ps@IEEEtitlepagestyle{%
 \old@ps@IEEEtitlepagestyle%
 \def\@oddhead{\strut\hfill#1\hfill\strut}%
 \def\@evenhead{\strut\hfill#1\hfill\strut}%
 }%
 \ps@headings%
 }
\makeatother

\confheader{%
Accepted for Presentation at IEEE Wireless Communications and Networking Conference (WCNC), 2024}

\begin{document}

\title{RIS-Enabled Integrated Sensing and Communication for 6G Systems}

\author{\IEEEauthorblockN{Dexin Wang\IEEEauthorrefmark{1}, Ahmad Bazzi\IEEEauthorrefmark{1}, and Marwa Chafii\IEEEauthorrefmark{1}\IEEEauthorrefmark{2}}
\IEEEauthorblockA{\IEEEauthorrefmark{1}Engineering Division, New York University (NYU), Abu Dhabi, UAE.}
	\IEEEauthorblockA{\IEEEauthorrefmark{2}NYU WIRELESS, NYU Tandon School of Engineering, New York, USA}}



\maketitle

\begin{abstract}
The following paper proposes a new target localization system design using an architecture based on reconfigurable intelligent surfaces (RISs) and passive radars (PRs) for integrated sensing and communications systems. 
The preamble of the communication signal is exploited in order to perform target sensing tasks, which involve detection and localization.
The RIS in this case can aid the PR in sensing targets that are otherwise not seen by the PR itself, due to the many obstacles encountered within the propagation channel.
Therefore, this work proposes a localization algorithm tailored for the integrated sensing and communications RIS-aided architecture, which is capable of uniquely positioning targets within the scene. 
The algorithm is capable of detecting the number of targets along with estimating the position of targets via angles and times of arrival.
Our simulation results demonstrate the performance of the localization method in terms of different localization and detection metrics and for increasing RIS sizes.
\end{abstract}


\section{Introduction}
\Ac{ISAC} has been proposed to be one of the key and challenging pillars for 6G \cite{chafii2023twelve}, where both information for sensing and communications can be carried by the same waveform. It is reasonable to believe that following the mass potential deployment of technologies such as haptic medicine, extended reality, holographic teleportation, terahertz bands, and Internet of things devices, a bandwidth-hungry world will be the main concern for 6G researchers and engineers \cite{saad2019vision}. ISAC, fortunately, allows for shared spectrum as well as improved efficiency in both processing power and hardware, which is one of the main ways to alleviate this limitation imposed on bandwidth. Conventionally, we can distinguish between three categories of ISAC: \textit{joint-design} \cite{10018908}, which compromises both sensing and communications; \textit{radar-centric} ISAC, which focuses on carrying information onto radar signals; while the last one, \textit{communication-centric} ISAC, the focus of this work, which aims at performing sensing using communication waveforms. Moreover, existing and widely used waveforms can also save costs on equipment, installation, and maintenance. 

Fortunately in communications, the preamble is such a widely used deterministic sequence. Hence, although they are intended to identify the cell identity of a \ac{BS} and synchronize the \ac{UEs} in the cell with the \ac{BS}, they are also strong candidates for target localization. It is also worth noticing that to perform target localization, having \ac{LoS} between the targets and the \ac{PR} is optimal but not necessarily guaranteed. However, \ac{RISs}, \cite{renzo2019ris}, which are large surfaces that can self-readjust its reflection properties and hence the channel properties as a whole, have drawn considerable attention in the past few years. By easily spreading over a large area, paths with LoS are more easily guaranteed.

There are myriad pieces of research that investigate RIS-assisted ISAC for sensing. However, some work such as \cite{wang2022dl} designs the reflection matrices, i.e. the settings of the phase shifters on the RIS, through channel estimation or \ac{UE} feedback. Moreover, in works such as \cite{kim2022monostatic}, the separation of paths from different targets is done by precoding instead of exploiting the spatial filtering properties of the RIS. In \cite{haider2023active}, the scheme depends on active elements which lead to increased power, noise, and complexity. In addition, in \cite{zuo2023noma} and \cite{he2022uav}, only one-dimensional estimation of the \acp{AoA} is done.

In this paper, we introduce a novel \ac{RIS}-enabled system model employing a \ac{PR}, as well as an \ac{AP} dedicated to communication-only tasks. The model accommodates a passive \ac{RIS} covering the entire view of the environment, which has the benefit of sensing reflections of the targets, then re-directing it back to the \ac{PR}. Indeed, due to obstacles, such as buildings, there may not be a \ac{LoS} component between the target and the \ac{PR}. Furthermore, we propose a novel target localization algorithm for joint \ac{ToA} and \ac{AoA} estimation of the targets, even when the number of targets is unknown, and even when no \ac{LoS} exists between the \ac{PR} and the target. The algorithm is also able to discriminate between targets located at the same direction, relative to the \ac{RIS}. Numerical simulations show that under very low \ac{SNR}, we can achieve a mean absolute error of about 30 cm in a 1000 m $\times$ 1000 m cell using 64 RIS elements.

The paper is organized in the following manner. Section \ref{sec:system-model} describes the system model, including the transmitted signal, the channel and received signal, the beamforming of the PR, and mapping the AoA-ToA pairs to Cartesian coordinates. Section \ref{sec:target-localization} describes the localization scheme for finding the AoA and ToA for each target through a step-by-step explanation. Section \ref{sec:simulation-results} describes the environment, test metrics, and test parameters used in simulating the system and outlines the simulation results. Section \ref{sec:conclusions-and-future-insights} outlines the conclusions of this work and proposes potential future directions.
 
\textbf{Notation}: Upper-case and lower-case boldface letters denote matrices and vectors, respectively. 
$(.)^T$, $(.)^*$, and $(.)^H$ represent the transpose, the conjugate and the transpose-conjugate operators. 
Furthermore, the set of all complex-valued $N \times M$ matrices is $\mathbb{C}^{N \times M}$. 
The $k^{th}$ entry of vector $\pmb{x}$ is denoted as $[\pmb{x}]_k$.
To index the $(i,j)^{th}$ entry of $\pmb{A}$, we use $[\pmb{A}]_{i,j}$.
Its $i^{th}$ row and $j^{th}$ column are indexed as $[\pmb{A}]_{i,:}$ and $[\pmb{A}]_{:,j}$, respectively.
Moreover, $\angle$ denotes an angle. 
Also, $\lbrace a : b \rbrace$ is used to denote the set $\{a,a+1,a+2,...,b\}$. 
The diagonal operator is $\diag$. 
The Kronecker product is $\otimes$.
An all-ones vector of size $N$ is denoted as $\pmb{1}_N$.
Also, $\argsmax_{\mathcal{C}}{f(x)}$ indicates finding the arguments that produce the local maxima of the function $f(x)$ subject to constraint $\mathcal{C}$.

\section{System Model}
\label{sec:system-model}
\begin{figure}[!t]
\centering
\includegraphics[width=3.5in]{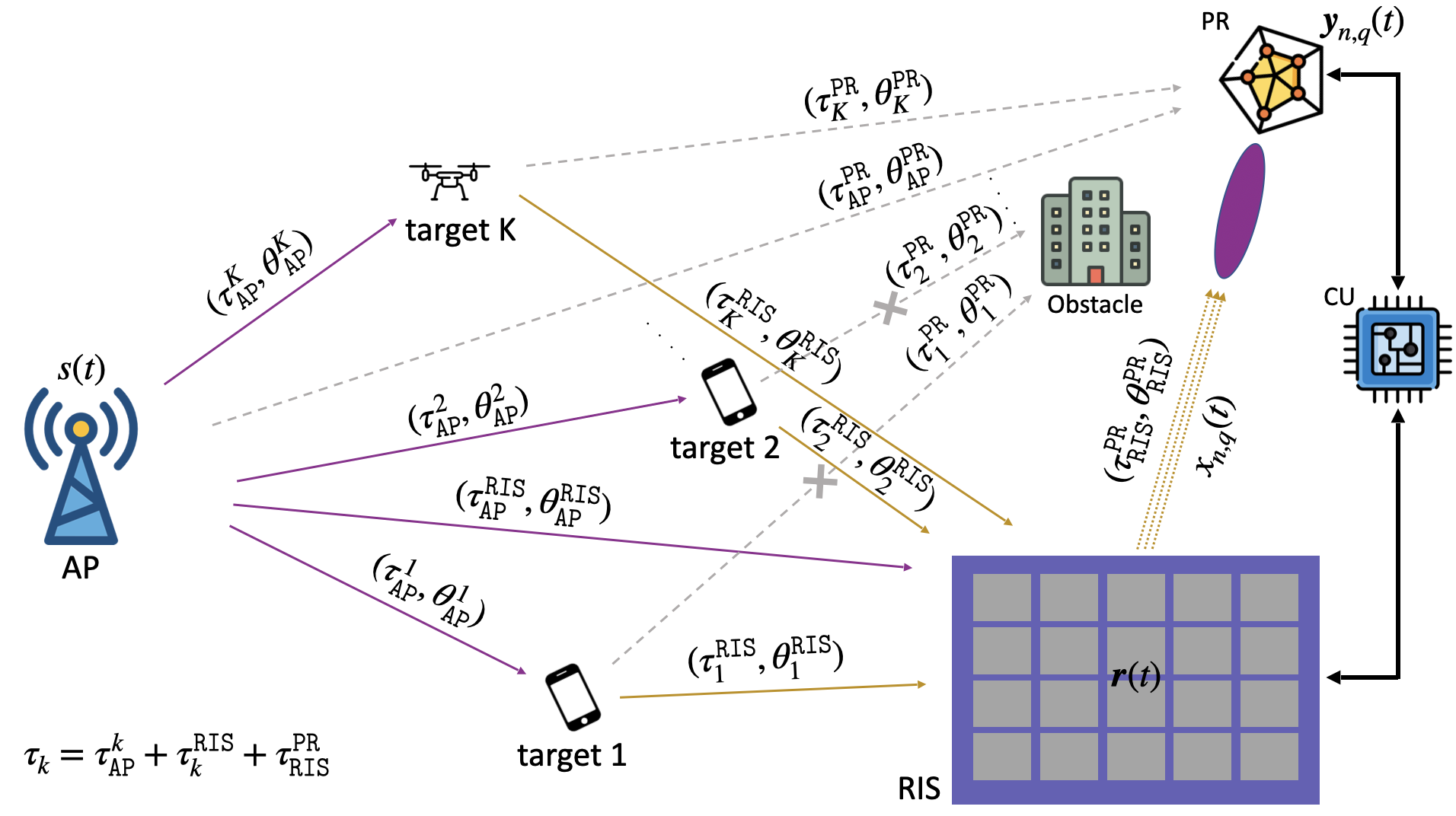}
\caption{Illustration of the scenario containing an AP, a RIS, a PR, $K$ targets, and potential obstacles.}
\label{System}
\end{figure}

We consider a scenario where the transmitted signal from the \ac{AP} hits $K$ targets, where $K$ is unknown. We focus on the part that bounces off from the targets toward the RIS, as well as the direct path from the AP to the RIS. Thanks to the reconfigurable behavior of the RIS, the received signal at the \ac{RIS} can be beamformed towards the PR. After the PR properly beamformers towards the RIS, it estimates the \ac{AoAs} ($\theta_k^{\tt{RIS}}$) and \ac{ToAs} ($\tau_k$) based on the received signal from the RIS. Then, the two-dimensional Cartesian coordinates of each target $k$, $\pmb{p}_k \in \mathbb{R}^{2\times 1}$, are determined from the two sensing parameters. We do not rely on the direct \ac{LoS} path between the AP and the PR and those between the $K$ targets and the PR. For those paths, there exists a higher possibility that LoS is not guaranteed, especially in dense environments such as cities or forests. On the other hand, the paths that pass through the RIS are less subject to blockage since RISs can easily span a large area such as the facades of buildings, \cite{kisseleff2020city}.

In this work, the system considers a rectangular cell with a specified width and height. One AP operating with existing communication standards such as 5G-NR, one PR with $N_{\tt{PR}}$ antenna elements, and one RIS with $M$ elements are specified with known $x$- and $y$- Cartesian coordinates $\pmb{p}_{\tt{AP}}, \pmb{p}_{\tt{RIS}}, \pmb{p}_{\tt{PR}} \in \mathbb{R}^{2\times 1}$, respectively. In Section \ref{subsec:channel-model}, we characterize the location of different nodes in terms of their \ac{AoA} and \ac{ToA} and in Section \ref{subsec:mapping}, we map these sensing parameters to positions $\pmb{p}$. To this end, our system model is depicted in Fig. \ref{System}.
We note that the \ac{RIS} and \ac{PR} can be connected to a \ac{CU}, in which all the computations and adjustments of \ac{RIS} phases are computed and controlled.

\subsection{Transmitted Signal}
Among the deterministic communication preambles, \ac{ZC} sequences are used in synchronization signals in 4G LTE and 5G NR, and they are repeated twice for each communication frame\cite{omri2019sync}. They are known for their \ac{CAZAC} property \cite{andrews2023zc}. The constant amplitude property ensures that \ac{ZC} sequences are less prone to distortion when passed through elements, such as amplifiers with nonlinear gain. At the same time, the zero cyclic auto-correlation indicates that when a \ac{ZC} sequence is correlated with its cyclically shifted version, the output remains below noise level except for the timestamp of the cyclic shift compared to the original sequence. Moreover, since covariance is a bi-linear operation, multiple peaks can be resolved for the sum of different cyclically shifted versions of the same ZC sequence. The transmitted signal hence takes the form of the \ac{ZC} sequence defined as below
\begin{equation}
	\label{eq:ZC}
    s_r[m]=e^{-j\pi r \frac{m(m+1)}{{N_{zc}}}} ,
\end{equation}
where ${N_{zc}} \in \{2m+1|m\in\mathbb{N}\}$ is the length of the sequence, $r \in \{1,2,3,...,{N_{zc}}-1\}$ is the root index and should be relatively prime with ${N_{zc}}$, and $m \in \{0,1,2,...,{N_{zc}}-1\}$ is the discrete time index, \cite{andrews2023zc}. During a supposed communication event, ZC sequences with the same specified length and root index are repeatedly transmitted from the AP.

\subsection{Channel and Received Signal}
\label{subsec:channel-model}
For simplicity, suppose that the signal $s(t)$ is generated from a \ac{ZC} sequence $s_r[m]$ using an ideal \ac{DAC}, the signal is transmitted from the \ac{AP} without beamforming, an $M$-element RIS is used, and there are $K$ targets, the received signal at the RIS is
\begin{equation}
	\label{eq:r-RIS}
    \pmb{r}(t) = \pmb{A}(\Theta_{1:K}^{\tt{RIS}})\pmb{s}(t)+\alpha_0\pmb{a}_M(\theta_{\tt{AP}}^{\tt{RIS}})s(t-\tau_{\tt{AP}}^{\tt{RIS}}) ,
\end{equation}
where $\pmb{A}(\Theta_{1:K}^{\tt{RIS}}) \in \mathbb{C}^{M \times K}$ is the steering matrix resulting from the steering vectors of each path between the targets and the RIS antennas manifold, namely
\begin{equation}
    \label{eq:steeringmat}
    \pmb{A}(\Theta_{1:K}^{\tt{RIS}}) = 
    \begin{bmatrix}
        \pmb{a}_M(\theta_1^{\tt{RIS}}) &
        \pmb{a}_M(\theta_2^{\tt{RIS}}) &
        ... &
        \pmb{a}_M(\theta_K^{\tt{RIS}})
    \end{bmatrix},
\end{equation}
where $\pmb{a}_M(\theta) \in \mathbb{C}^{M \times 1}$ is the steering vector of the $M$-element RIS, and $\theta_i^j$ is the AoA at $j$ from $i$, where $i$ and $j$ are either the AP, RIS, PR, or targets. On the other hand, $\pmb{s}(t)$ contains the delayed and scaled versions of $s(t)$, where its $k^{th}$ entry is
\begin{equation}
    \label{eq:smatrix}
    [\pmb{s}(t)]_k =  
    \alpha_k s(t-\tau_{\tt{AP}}^k-\tau_k^{\tt{RIS}})
    \in \mathbb{C},
\end{equation}
where $\alpha_k$ is the channel gain of the path via target $k$. Also, $\tau_i^j$ is the delay from node $i$ to $j$. Note that $i$ and $j$ are either the AP, RIS, PR, or targets. The reflected signal from RIS is
\begin{equation}
	\label{eq:x-RIS}
    x_{n,q}(t) = 
    \pmb{a}_{M}^T(\phi_{\tt{RIS}}^{\tt{PR}})
    \diag(\pmb{v}_{n,q}) \pmb{r}(t) ,
\end{equation}
where $\pmb{v}_{n,q} \in \mathbb{C}^{M \times 1}$ is the phase shifts due to the reflecting elements of the RIS at the $n^{th}$ epoch within the $q^{th}$ phase and $\phi_{\tt{RIS}}^{\tt{PR}}$ is the \ac{AoD} from \ac{RIS} to \ac{PR}. 
We define an epoch as a collection of samples of $\pmb{r}(t)$ in \eqref{eq:r-RIS}, and we define a phase as a collection of epochs.
The received signal at the PR at epoch $n$ and phase $q$ is
\begin{equation}
{\small
	\label{eq:yn-PR}
    \begin{split}
    \pmb{y}_{n,q}(t)
    & = \rho_{\tt{RIS}}^{\tt{PR}}\pmb{a}(\theta_{\tt{RIS}}^{\tt{PR}})x_{n,q}(t-\tau_{\tt{RIS}}^{\tt{PR}})  + B_{\tt{AP}}\rho_{\tt{AP}}^{\tt{PR}}\pmb{a}(\theta_{\tt{AP}}^{\tt{PR}})s(t-\tau_{\tt{AP}}^{\tt{PR}}) \\
    & + \sum\nolimits_{k=1}^K B_k\rho_k\pmb{a}(\theta_{k}^{\tt{PR}})s(t-\tau_{\tt{AP}}^k-\tau_k^{\tt{PR}})+\pmb{\epsilon}_{n,q}(t) , \\
    \end{split}
    }
\end{equation}
where $\pmb{y}_{n,q}(t) \in \mathbb{C}^{N_{\tt{PR}} \times 1}$. 
The first three terms correspond to the information reflected from the RIS, direct LoS propagation from the AP, and bounce-off from the targets. 
The last term $\pmb{\epsilon}_{n,q}(t)$ is \ac{AWGN}. The terms $B_{\tt{AP}}$ and $B_k$ are binary random variables that equal 0 when the LoS is blocked and 1 otherwise. Sampling $\pmb{y}_{n,q}(t)$ at $L$ time instances, we have the sampled data matrix
\begin{equation}
    \label{eq:Yn}
    \pmb{Y}_{n,q} =
    \begin{bmatrix}
        \pmb{y}_{n,q}(1)& \pmb{y}_{n,q}(2)&...&\pmb{y}_{n,q}(L)
    \end{bmatrix}
    \in \mathbb{C}^{N_{\tt{PR}} \times L} .
\end{equation}
Moreover, the \ac{PR} collects $\pmb{Y}_{n,q}$ in \eqref{eq:Yn} and beamforms as 
\begin{equation}
    \label{eq:Z}
   [ \pmb{Z}(\pmb{V}_q) ]_{n,:} =
   \pmb{w}^H \pmb{Y}_{n,q}
    \in \mathbb{C}^{1 \times L}, \ \forall n,q,
\end{equation}
where $\pmb{Z}$ is a function of $\pmb{V}_q = [\pmb{v}_{1,q} \ldots \pmb{v}_{N_{\text{epoch}},q} ]$, the reflection matrix of the RIS given the time slot $q \in \{0:\hat{K}_\theta\}$, where $\hat{K}_\theta$ is the number of estimated directions. The details of how $\pmb{V}_t$ changes is further outlined in Section \ref{subsec:findingAoAs}. Then, from equation \eqref{eq:yn-PR}, the useful part of the received signal looking at the direction of $\theta_{\tt{RIS}}^{\tt{PR}}$ can be represented as $\rho_{\tt{RIS}}^{\tt{PR}} \pmb{w}^H \pmb{a}(\theta_{\tt{RIS}}^{\tt{PR}})x_{n,q}$ whose power can be maximized, under a norm-constraint, by adjusting \cite{bazzi2022rispr}
\begin{equation}
    \label{eq:w}
    \pmb{w} =\left\lVert \pmb{a}(\theta_{\tt{RIS}}^{\tt{PR}})\right\rVert^{-2} \pmb{a} (\theta_{\tt{RIS}}^{\tt{PR}}).
\end{equation}

\begin{figure}[!t]
\centering
\includegraphics[width=3.25in]{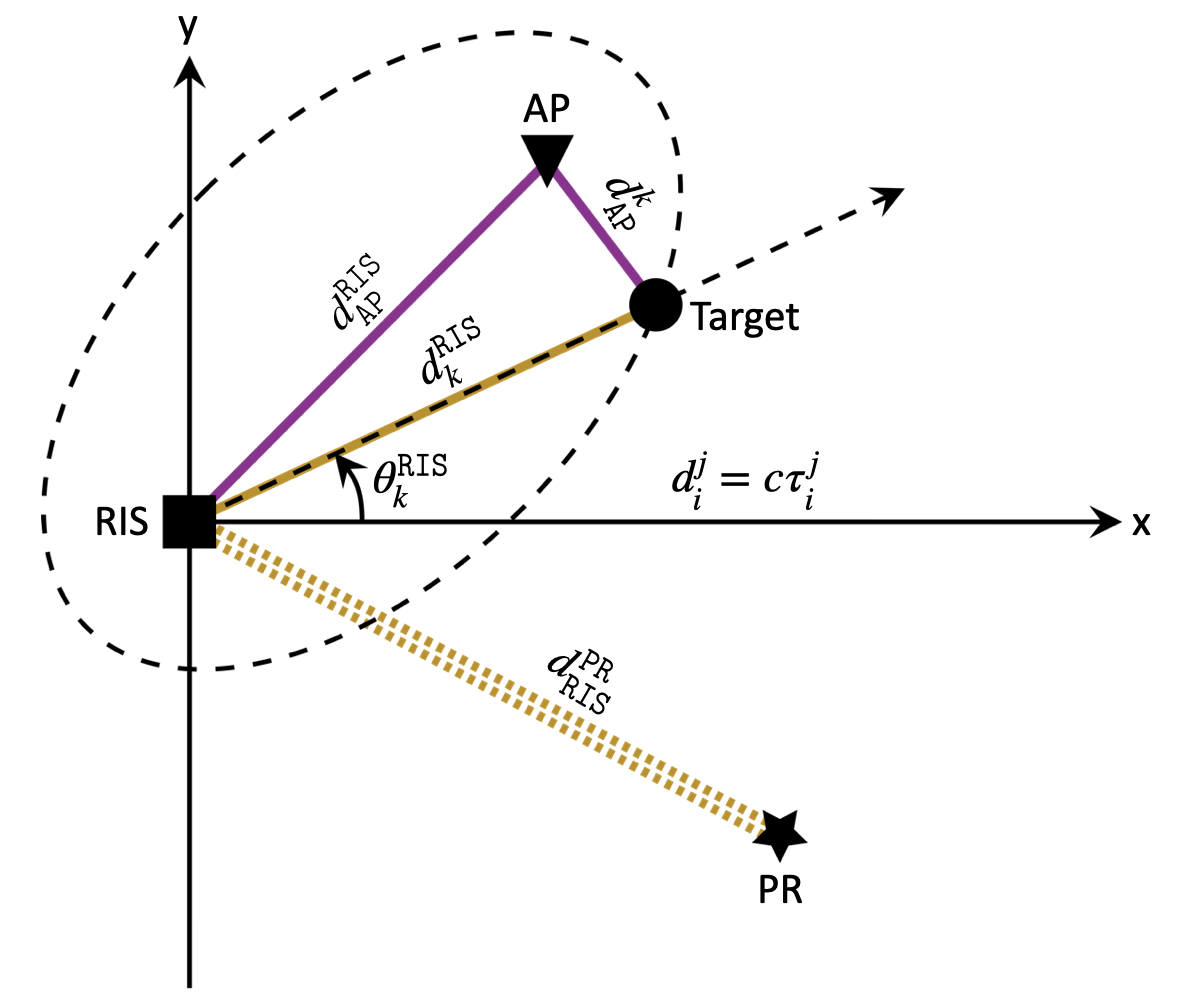}
\caption{The problem of mapping $(\theta_k^{\tt{RIS}},\tau_k)$ to $\pmb{p}_k$ can be modeled as finding an intersection of a secant line through an ellipse given that the target is not in the LoS between the AP and RIS. The mapping is bijective.}
\label{Mapping}
\end{figure}

\subsection{Mapping $(\theta_k^{\tt{RIS}},\tau_k)$ to $\pmb{p}_k$}
\label{subsec:mapping}
Once $\theta_k^{\tt{RIS}}$ and $\tau_k$ are determined, the problem of finding $\pmb{p}_k$ is equivalent to finding an intersection of a secant line through an ellipse given that the target is not in the LoS between the AP and RIS. The scenario is depicted in Fig. \ref{Mapping}. First, it is easy to see that
\begin{equation}
    \label{eq:tau-d}
    d_{\tt{AP}}^{k} + d_{k}^{\tt{RIS}} + d_{\tt{RIS}}^{\tt{PR}} = c \tau_k,
\end{equation}
where $i$ and $j$ are either the AP, RIS, PR, or targets, $d_i^j$ is the distance between $i$ and $j$, and $c$ is the speed of light, the propagation velocity of the signal. Since the positions of the AP, RIS, and PR are known, $d_{\tt{AP}}^{\tt{RIS}}$ and $d_{\tt{RIS}}^{\tt{PR}}$ can be found by the Pythagorean theorem $d_i^j = \left\lVert \pmb{p}_i-\pmb{p}_j \right\rVert$.
As we now know $\pmb{p}_{\tt{AP}}$, $\pmb{p}_{\tt{RIS}}$, and $d_{\tt{AP}}^{k} + d_{k}^{\tt{RIS}}$, we can conclude that the target lies on the ellipse whose foci are the AP and the RIS, whose semi-major axis is $a = (d_{\tt{AP}}^{k} + d_{k}^{\tt{RIS}})/2$ and whose semi-minor axis is $b = \sqrt{a^2-d_{\tt{AP}}^{\tt{RIS}}/2}$. Additionally, we can easily see that the target also lies on the line in the direction $\theta_k^{\tt{RIS}}$. Next, we first consider the case when the ellipse is not tilted and centered at the origin. Here, the line can be written as $y = px+q$, where $p$ and $q$ are the slope and y-intercept, respectively, and the ellipse can be written in its standard equation $\frac{x^2}{a^2}+\frac{y^2}{b^2}=1$. Plugging the ellipse relation into the line one, we can solve the quadratic equation
\begin{equation}
    \label{eq:lineellipse}
    (a^2p^2+b^2)x^2+2a^2pqx+a^2(q^2-b^2)=0,
\end{equation}
to deduce the coordinates under this reference frame. 
Performing a rotation by $\theta^{\tt{RIS}}_{\tt{AP}}$ and translation via $(\pmb{p}_{\tt{AP}}+\pmb{p}_{\tt{RIS}})/2$, the center of the ellipse, we can recover the absolute Cartesian coordinates. Usually, there exist two solutions for this kind of problem. Nonetheless, since we specify that $\theta_k^{\tt{RIS}} \in (-90^\circ,90^\circ)$, the solution on the left side of the RIS can be easily eliminated and $\pmb{p}_k$ is hence the other one. Moreover, it is obvious that such a map is bijective, meaning that the one $(\theta_k^{\tt{RIS}},\tau_k)$ pair corresponds uniquely to a $\pmb{p}_k$ and vice versa.
We can now state our problem: Based on $\pmb{Y}_{n,q}$, estimate the target locations $\pmb{p}_1 \ldots \pmb{p}_K$, where $K$ is unknown.

\section{Target Localization}
\label{sec:target-localization}
\begin{figure}[!t]
\centering
\includegraphics[width=3.5in]{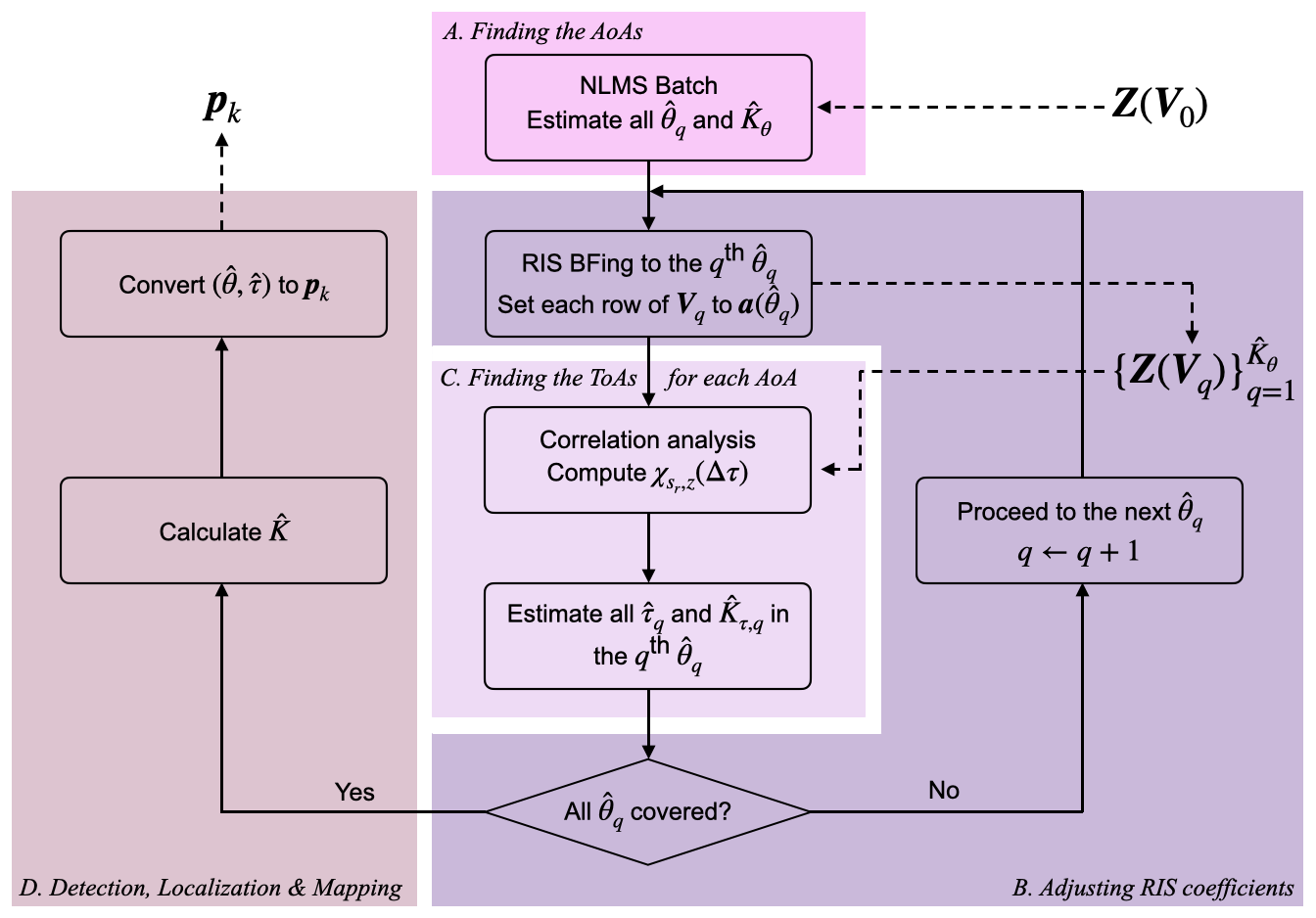}
\caption{The target localization scheme can be summarized into the four parts with different colors. The inputs to and outputs from the system are shown in dotted lines while the process is shown in regular lines.}
\label{FlowChart_New}
\end{figure}

The target localization process is illustrated in Fig. \ref{FlowChart_New}. The AoAs are estimated through an NLMS batch algorithm \cite{bazzi2022rispr}, and the ToAs are derived from analyzing the cross-correlation between the transmitted and received sequences. To associate each target with a unique AoA-ToA pair, the ToA analysis is performed for each estimated AoA. This is accomplished by changing the RIS reflection properties. 
For this reason, the initial phase, i.e. $q=0$, performs \ac{AoA}-only estimation. Then, the $q^{th}$ phase performs \ac{ToA} estimation per estimated \ac{AoA}.

\subsection{Finding the AoAs (Phase $q=0$)}
Initially, our goal is to maximize the power collected from the $K$ targets at the RIS and minimize the power of all other directions, which in this case is due to \ac{AP}. 
We optimize for the aforementioned criterion while maintaining the constraint of unity gain ($ \vert [\pmb{V}_0]_{i,j} \vert  = 1, \forall i,j$) for the passive RIS. The solution of the unconstrained version of the optimization problem from our work \cite{bazzi2022rispr} is
\begin{equation}
    \label{eq:Vbar}
    \overline{\pmb{V}}_0=\mathcal{P}_{\tilde{\pmb{a}}(\theta_{\tt{AP}}^{\tt{RIS}})}^{\perp}=\pmb{I}_{M}-\left\|\tilde{\pmb{a}}(\theta_{\tt{AP}}^{\tt{RIS}})\right\|^{-2} \tilde{\pmb{a}}(\theta_{\tt{AP}}^{\tt{RIS}}) \tilde{\pmb{a}}^{H}(\theta_{\tt{\tt{AP}}}^{\tt{\tt{RIS}}}),
\end{equation}
and to satisfy the constant modulus constraint, we calculate our initial RIS reflection matrix as
\begin{equation}
    \label{eq:V}
    \pmb{V}_0 = \exp\{{j \angle\mathcal{P}_{\tilde{\pmb{a}}(\theta_{\tt{AP}}^{\tt{RIS}})}^{\perp} \pmb{\Gamma}}\}.
\end{equation}
where $\pmb{\Gamma} \in \mathbb{C}^{M \times N_{epoch}}$ is a random matrix drawn from a standard Gaussian distribution and $\tilde{\pmb{a}}(\theta_{\tt{AP}}^{\tt{RIS}}) = \diag[\pmb{a}_{M}(\phi_{\tt{RIS}}^{\tt{PR}})]\pmb{a}(\theta_{\tt{AP}}^{\tt{RIS}})$. The AoAs $\hat{\theta}_k^{\tt{RIS}}$ and the number of distinct directions $\hat{K}_\theta$ are then estimated\footnote{As an example, if $2$ targets are present at the same angle relative to \ac{RIS}, then $K_\theta = 1$, otherwise $K_\theta = 2$.} through the NLMS batch algorithm described in our previous work \cite{bazzi2022rispr}.
Note that the true value $K_\theta$ is upper-bounded by $K$. We distinguish between two extreme cases: \textit{(i)} all targets reside in the same angle $\theta_1^{\tt{RIS}} = \ldots = \theta_K^{\tt{RIS}} = \theta^{\tt{RIS}}$, which means $K_\theta = 1$, and \textit{(ii)} when all the \acp{AoA} are distinct, i.e. $\theta_k^{\tt{RIS}} \neq \theta_{k'}^{\tt{RIS}}$ for all $k \neq k'$. In this case, $K_\theta = K$. As we shall proceed, we rely on the temporal aspect of the \ac{ZC} sequence in order to separate the targets in time domain.

\subsection{Adjusting \ac{RIS} coefficients (Phase $q>0$)}
Using the solution given by \eqref{eq:w}, but this time at the \ac{RIS}, spatial filtering towards each estimated direction is accomplished by beamforming the RIS towards that direction by changing each row of the RIS reflection matrix $\pmb{V}_q$, to $\pmb{a}(\hat{\theta}_q^{\tt{RIS}})$ for all $q>0$, i.e.
\begin{equation}
    \pmb{V}_q
    =
     \diag[\pmb{a}_{M}(\phi_{\tt{RIS}}^{\tt{PR}})]
     \big(
    \pmb{a}^T(\hat{\theta}_q^{\tt{RIS}})
    \otimes
    \pmb{1}_{N_{epoch}}
    \big).
\end{equation}
This allows the \ac{RIS} to \textit{select} targets that are located at direction $\hat{\theta}_q^{\tt{RIS}}$ so as to estimate each of their \acp{ToA}, separately.




\subsection{Finding the ToAs for each AoA (Phase $q>0$)}
\label{subsec:findingAoAs}
\paragraph{Correlation Analysis}
Exploiting the \ac{CAZAC} property of ZC sequences and the bi-linearity of the covariance operation, we employ a correlator to deduce the ToAs. To consider all epochs, the test sequence is the average of $\pmb{Z}$ over all rows, namely
\begin{equation}
    \label{eq:z}
    \pmb{z} = \frac{1}{N_{epoch}}\sum\nolimits_{i = 1}^{N_{epoch}}[\pmb{Z}(\pmb{V}_q)]_{i,:},
\end{equation}
where $n \in \{0,1,2,...,L-1\}$. Next, suppose that $N_{zc} = L$ and an ideal \ac{ADC} is used so that the PR and AP are perfectly synchronized with no sampling frequency ($f_{samp}$) or phase offsets, the magnitude of the discrete cross-correlation between the original and test sequences can be calculated by
\begin{equation}
    \label{eq:corr}
    \chi_{s_r,z}(\Delta\tau) = \left\lvert \sum\nolimits^{L/2}_{n=-L/2}{s_r[n]
    z_c^*[n-\Delta\tau}] \right\rvert,
\end{equation}
where $\Delta\tau$ is the correlation shift in number of samples and is related to $\tau$ by a factor of $f_{samp}^{-1}$. Furthermore, $z_c$ is a centered time-series generated simply by centering $\pmb{z}$.

\paragraph{Estimate ToAs and Number of Targets residing in $\hat{\theta}_q^{\tt{RIS}}$} 
Thanks to the association scheme, the correlator performs a correlation analysis for each estimated direction. The \acp{ToA} are determined from a peak-finding search of the normalized correlation output of that direction. We define a threshold, say $g_\tau$, for determining if a peak qualifies as an estimated target. For the $q^{th}$ estimated direction, the \acp{ToA} can be estimated as
\begin{equation}
    \label{eq:tauk}
    \hat{\tau}_{q,i}^{\tt{RIS}} \in \argsmax_{\overline{\chi}_{s_r,z}>g_\tau }{[\chi_{s_r,z}(\Delta\tau)]}, \quad 
    i = 1 \ldots \hat{K}_{\tau,q},
\end{equation}
where $\overline{\chi}_{s_r,z}$ is the normalized ${\chi}_{s_r,z}$. The number of targets estimated in direction $\hat{\theta}_q^{\tt{RIS}}$ is hence the number of all qualifying peaks, and it is denoted as $\hat{K}_{\tau,q}$. Therefore, the targets residing in angle $\hat{\theta}_q^{\tt{RIS}}$ are $\hat{\tau}_{q,1}^{\tt{RIS}} \ldots \hat{\tau}_{q,\hat{K}_{\tau,q}}^{\tt{RIS}} $.
To this end, steps $B$ and $C$ are repeated until $q$ covers all $\hat{K}_\theta$ directions.

\subsection{Detection, Localization \& Mapping}
Once we have all the estimated \ac{AoA}/\ac{ToA} pairs, we proceed to the last step, herein, in order to detect the number of targets, then perform localization by mapping their \ac{AoA}/\ac{ToA} pairs to a Cartesian coordinate. Regarding detection, the total number of targets is estimated by summing the number of targets in each estimated direction, namely
\begin{equation}
    \label{eq:khat}
    \hat{K} = \sum\nolimits_{q = 1}^{\hat{K}_\theta} \hat{K}_{\tau,q}.
\end{equation}
Moreover, the Cartesian coordinates of the targets can be found from the $(\hat{\theta},\hat{\tau})$ pairs through the mapping mechanism described in Section \ref{subsec:mapping}.

\section{Simulation Results}
\label{sec:simulation-results}
\begin{figure}[!t]
\centering
\includegraphics[width=3in]{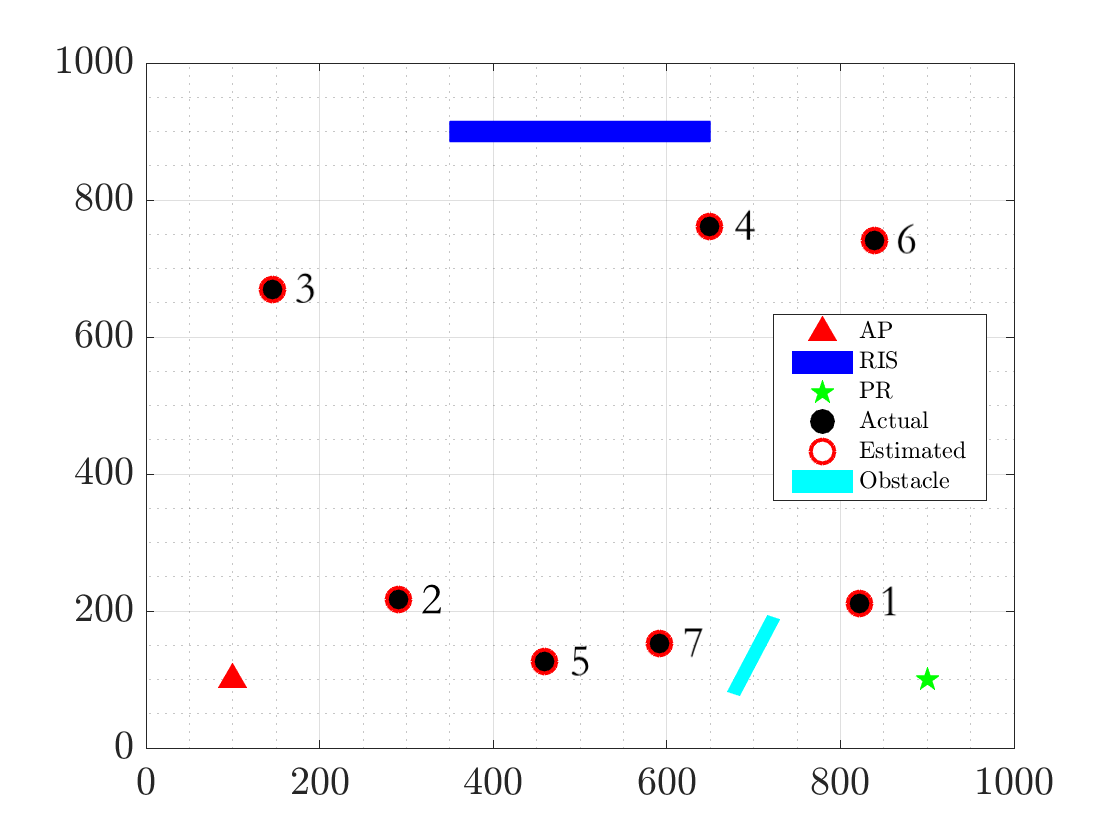}
\caption{Visualization of simulation of the target localization algorithm in a $1000 \text{ m} \times 1000 \text{ m}$ rectangular cell.}
\label{Grid}
\end{figure}

Fig. \ref{Grid} depicts the wireless environment, where multiple targets (black dots) are randomly placed in a $1000 \text{m} \times 1000 \text{m}$ rectangular cell. The \ac{AP}, \ac{RIS}, and \ac{PR} are placed in a way that the obstacle blocks the \ac{LoS} between three targets and the \ac{PR}. The number of antennas at PR is $16$; the used ZC index is $7$; and the length of the used ZC sequence is $1989$. Through exhaustive simulations, we find that the values of both 
 \ac{ToA} and \ac{AoA} thresholds are $0.3$.

\subsection{Test Metrics}
\vspace{-0.1cm}
\paragraph{Mean Squared Error (MSE)}
It is the average squared distances between the estimated targets and the corresponding actual targets. This is a factor only related to estimation. Since we only consider the targets that are actually estimated, and the estimated number of targets, $\hat{K}$, may be above or below the actual number of targets, $K$, we introduce the factor
\begin{equation}
    U = \min(K,\hat{K}).
\end{equation}
We then choose the $U$ pairs of actual and estimated targets that have the lowest distances for $\MSE$ computation, which is
\begin{equation}
    \MSE = \frac{1}{PU}\sum\nolimits_{p=1}^{P}\sum\nolimits_{u=1}^{U}[(x_u-\widehat{x}_u)^2+(y_u-\widehat{y}_u)^2],
\end{equation}
where $P$ is the number of Monte-Carlo simulations. 

\paragraph{Probability of Detection, $P_D$}
In order to show how well our target detector performs, we study the probability of detection, which is defined as 
\begin{equation}
    P_D = \frac{\text{no. of trials s.t. } K = \hat{K}}{\text{total no. of trials}}.
\end{equation}

\paragraph{Successful Recovery Probability (SRP)}
It is a factor related to both detection and estimation and is defined by
\begin{equation}
    \SRP = \frac{\text{no. of trials s.t. } \left\lvert\lvert \pmb{p}_k - \hat{\pmb{p}}_k \right\rvert\rvert \le \epsilon_{\SRP}}{\text{no. of trials}},
\end{equation}
where $\epsilon_{\SRP}$ is the error allowed for an estimation to be counted as a successful recovery of the original target coordinates. In this $1000 \text{ m} \times 1000 \text{ m}$ cell case, we set $\epsilon_{\SRP} =1 \text{ m}$.

\begin{figure}[!t]
\centering
\includegraphics[width=3in]{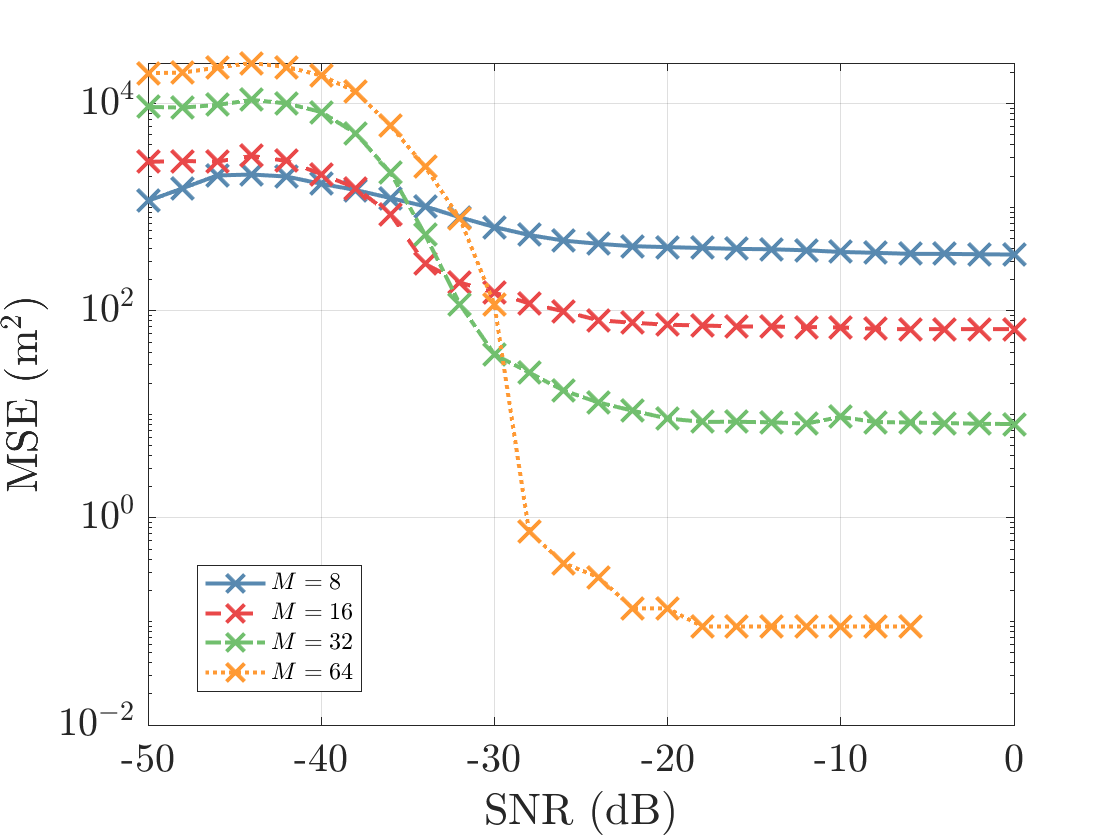}
\caption{\ac{MSE} performance vs. \ac{SNR} of $K = 2$ targets set in the scene by varying the number of \ac{RIS} elements $M$.}
\label{MSE_SNR}
\end{figure}


\subsection{MSE analysis as a function of number of \ac{RIS} elements}
In Fig. \ref{MSE_SNR}, we plot the \ac{MSE} performance as a function of \ac{SNR} for different values of \ac{RIS} elements, $M$.
For a fixed $M$, we observe that the \ac{MSE} always decays with increasing \ac{SNR}.
The rate of decrease is observed to be comparatively drastic between $\SNR = -40 \dB$ and $\SNR  = -20 \dB$. 
As we increase the number of \ac{RIS} elements, we see a significant drop in \ac{MSE} at $\SNR \geq -30 \dB$. 
The $M = 64$ case can achieve \ac{MSE} floor of about $0.089$ m$^2$, indicating a mean absolute error at about $30$ cm.
In addition, at a fixed desired \ac{MSE} of $5$ m$^2$, we can see that doubling the \ac{RIS} elements from $32$ to $64$ contributes to a gain of about $8\dB$.
This means that more \ac{RIS} elements improve the target location accuracy, as can be observed via \ac{MSE}, as well as mean absolute error, improvements.

\begin{figure}[!t]
\centering
\includegraphics[width=3in]{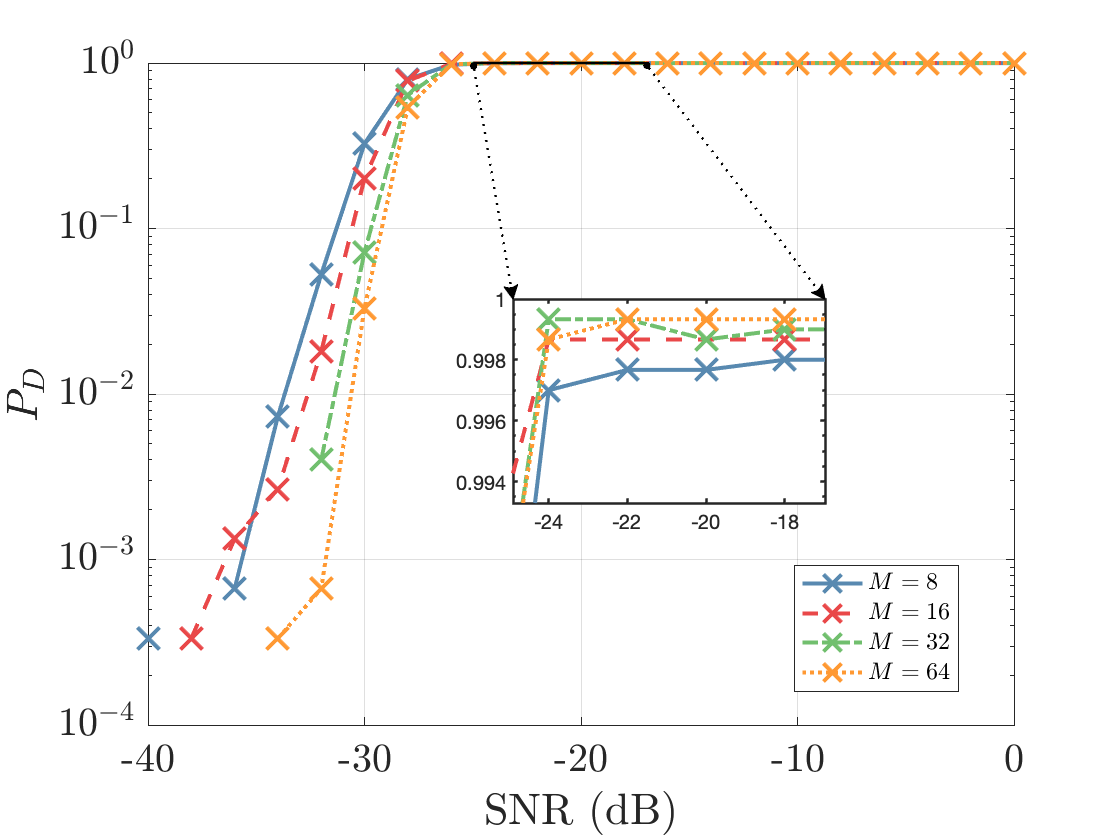}
\caption{Probability of detection performance vs. \ac{SNR} for fixed $K = 2$ targets set in the scene by varying the number of \ac{RIS} elements $M$.}
\label{PD_SNR}
\end{figure}
\begin{figure}[!t]

\centering
\includegraphics[width=3in]{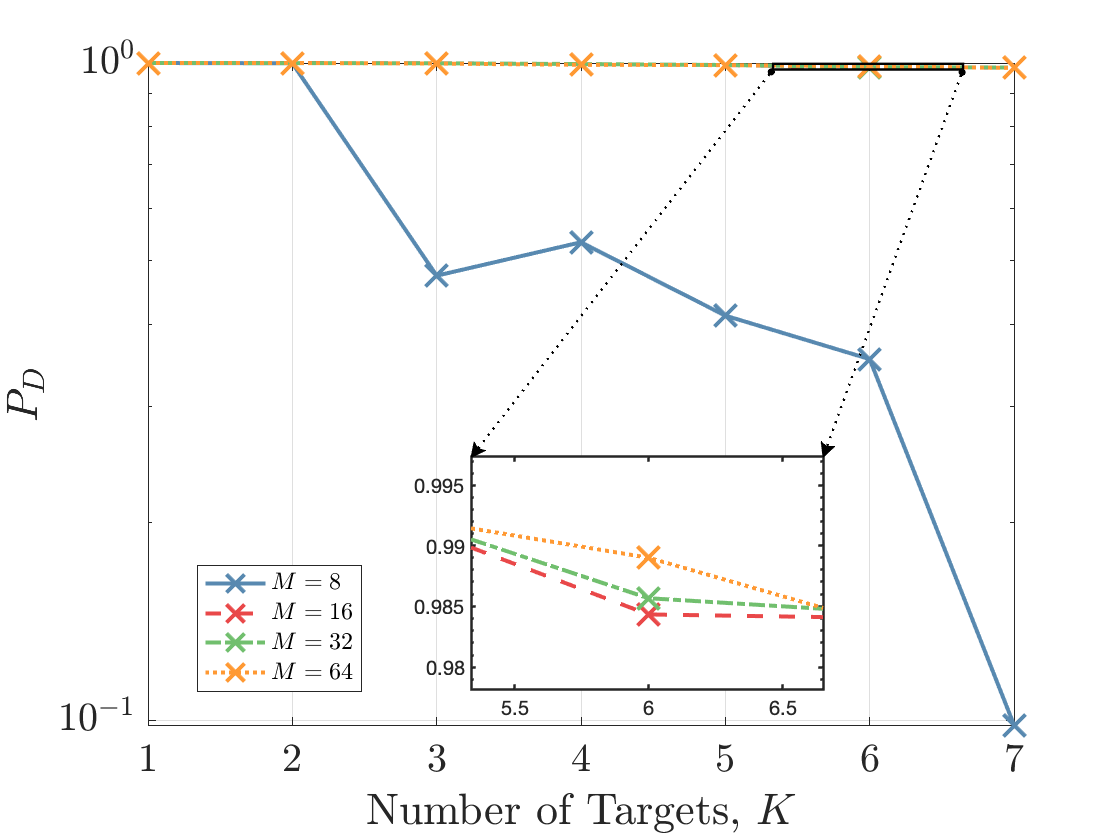}
\caption{$P_D$ vs. $K$, both thresholds for AoAs and ToAs are 0.3. The SNRs at both the RIS and the PR are set as $50\dB$.}
\label{PD_K}
\end{figure}

\subsection{Probability of detection gains}
In Fig. \ref{PD_SNR}, as \ac{SNR} increases, the probability of detection drastically grows and approaches 1 at around $-25 \dB$. 
Upon trespassing this \ac{SNR}, the probability of detection stabilizes around $1$, indicating successful detection for nearly all Monte-Carlo simulations at those \ac {SNR} levels.
Setting a desired probability of detection level to $P_D = 0.8$, we can observe a gain of about $1.5\dB$, when comparing $M=8$ with $M=64$.

In Fig. \ref{PD_K}, we study the detection probability as a function of the number of targets found in the scene. 
We observe that the probability of detection exceeds $0.95$ when $1 \leq K \leq 7$ starting from $M = 16$ \ac{RIS} elements, whereas the $M=8$ case can achieve this resolution only when $K \leq 2$ targets.
It can be concluded that the number of RIS elements is positively correlated with the detection capabilities of the overall proposed \ac{RIS}-aided architecture.

\begin{figure}[!t]
\centering
\includegraphics[width=3in]{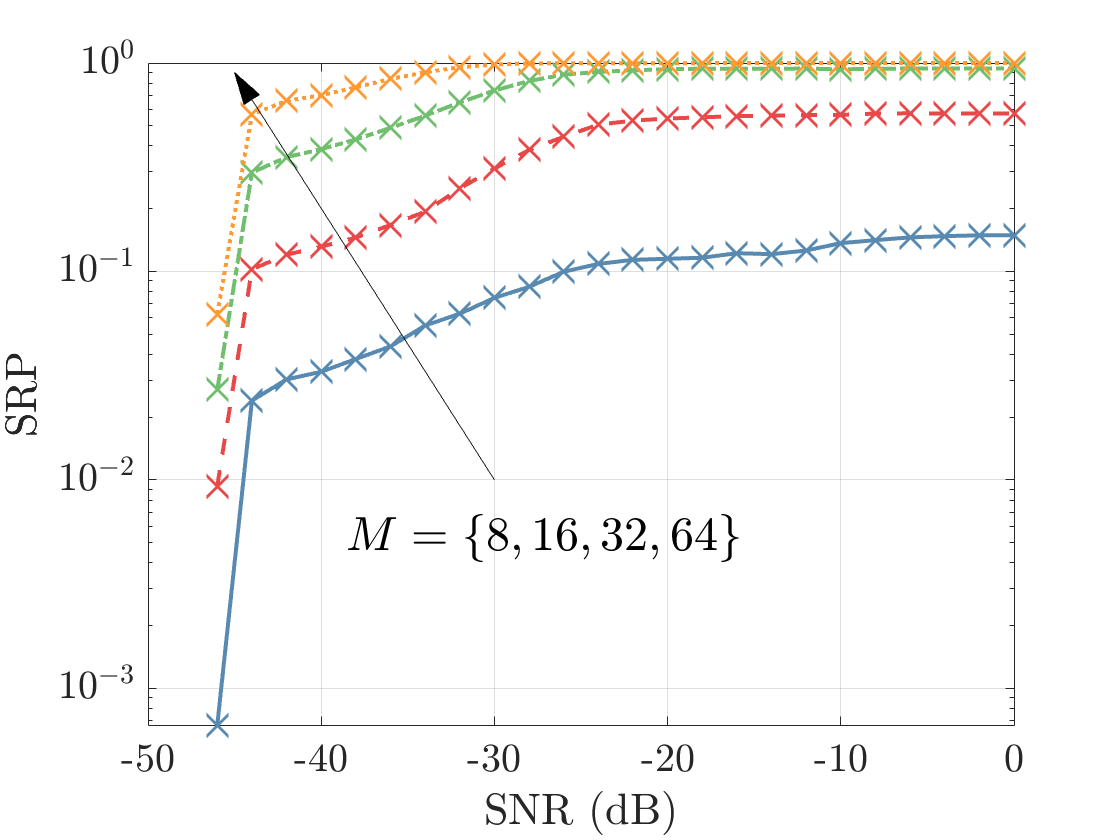}
\caption{SRP vs. SNR, two targets are deployed in the scene and both thresholds for AoAs and ToAs are 0.3 The value of the chosen $\epsilon_{\SRP}$ is 1 m.}
\label{SRP_SNR}
\end{figure}
\subsection{SRP gains with increasing $M$}
\vspace{-0.1cm}
In Fig. \ref{SRP_SNR}, we can see the positive impact of increasing the number of \ac{RIS} elements from an SRP perspective. 
At $\SNR \geq -20 \dB$, the $M = 64$ case gives SRP approaching 1, the $M = 32$ case achieves SRP up to around $0.9$, while the $M = 16$ case has an \ac{SRP} cap at around $0.6$. Moreover, for the $M = 8$ case, the \ac{SRP} is below $0.2$ and its trend does not have an observable ceiling.
This can be explained by the fact that more \ac{RIS} elements can offer better resolution capabilities and a better view of the environment so as to maximize the overall recovery probability of the system.

\section{Conclusions and Future Insights}
\label{sec:conclusions-and-future-insights}
In this paper, we propose a new \ac{RIS}-\ac{PR} based architecture for target localization in \ac{ISAC} for future 6G systems, which exploits already available signals used as part of a communication standard, like \ac{ZC} sequences. 
Indeed, increasing the number of \ac{RIS} elements has a favorable impact on localization accuracy, and target detection, as we can observe from the simulation results when utilizing the proposed localization method for simultaneous detection and localization. 
To further develop the comprehensiveness of this scheme, Doppler shift estimation can be investigated, which is interesting for target tracking applications. 
Moreover, more sophisticated algorithms should be aware of radio frequency impairments \cite{10061453}, such as carrier frequency offset and sampling time offset. 
Next, we consider improving the accuracy and coverage by using multiple \acp{RIS} and/or \acp{AP}. For instance, this prevents the inability to solve the case when the target is in the LoS between the AP and the RIS. Different scenes such as 3D environments and various channels are also considered as part of future work. 

\vspace{-0.1cm}
\bibliographystyle{IEEEtran}
\bibliography{refs.bib}

\vspace{12pt}

\end{document}